# Community Structure and Metacommunity Dynamics of Aquatic Invertebrates: a Test of the Neutral Theory

Michael M. Fuller[*] (1), Tamara N. Romanuk (2), and Jurek Kolasa (3)

*(1) Dept. of Biology, University of New Mexico, Albuquerque, NM, USA*

*(2) Département des sciences biologiques, Université du Québec, Montréal, QC, Canada*

*(3) Dept. of Biology, McMaster University, Hamilton, Ontario, Canada*

We used a metacommunity of 49 discrete communities of aquatic invertebrates to analyze the dynamical relationship between community and metacommunity species distributions as a test of the neutral theory of biodiversity and biogeography. At the community scale, observed variation in species richness and relative abundance was greater than predicted by neutral models, and revealed important differences among species in competitive ability and tolerance for predation. At the metacommunity scale, species with metacommunity proportions of less than 0.01% (38% of the observed metacommunity) were consistently more abundant than predicted by models. Our results are at variance with the neutral theory, and suggest that the use of an identical survival probability for all species in neutral models misrepresents substantial aspects of community assembly. Nevertheless, building and testing neutral models can provide valuable insights into the processes that determine species distributions.

---

[*] Email: mmfuller@unm,edu

**INTRODUCTION**

Despite many decades of theoretical development, experimentation, and debate, ecologists continue to disagree about whether species communities represent a select group of ecologically compatible organisms, or are simply a random subset of the regional species pool. Many studies have shown that competition and environmental conditions can select the species that can coexist at a particular location, and influence their local abundance [1-7]. These studies support the niche-assembly view of community structure, which is that communities represent non-random subsets of compatible species. This view is challenged by statistical hypotheses that propose that communities are random assemblages of species [8, 9]. Proponents of statistical hypotheses assert an absence of adequate evidence to support niche-assembly. Others counter that competition theory and observed regular patterns of association in species assemblages contradict statistical hypotheses. This disagreement has caused considerable controversy [10-17]. Recently the debate over community assembly has been invigorated by the development of a new theory called the neutral theory of biodiversity and biogeography [18-20].

The neutral theory is conceptually similar to the neutral theory of molecular evolution [21-23]. In the ecological neutral theory, the distribution of species in communities is governed by the random replacement of individuals of one species with those of another. The neutral theory assumes that species have identical per capita probabilities of birth, death, and migration. As such, species have an equal probability of winning a competitive interaction. In the neutral theory, the long term average of the numerical proportion of each species in a community that receives immigrants is equal to its proportional abundance in the metacommunity. Variation in species proportions among communities is due solely to the effects of finite community size (see below) and stochastic demographic processes. At the metacommunity scale, random demographic change slowly modifies the abundance of species. Those with very large populations change very slowly, while species with very small populations have a higher probability of loss by extinction and replacement by speciation. At the community scale, neutral models reproduce several well-known species distributions, including Taylor power laws, species-area relations, and the distribution of species abundance within communities [20, 24]. Furthermore, the distribution of species in certain types of communities, tropical forests in particular, is largely in accordance with the neutral theory [25, 26].

Although several workers have contributed to the neutral theory [20, 25, 27, 28], Hubbell (2001) provided the most complete development of the theory and related models. Hereafter, we mean specifically his approach when we refer to neutral models. These models describe the occurrence and abundance of species in a community in terms of the binomial distribution. Calculating the expected abundance of species for communities of more than a few individuals requires solving matrices of conditional probability equations, for which analytical solutions are intractable. However, community and metacommunity species distributions are well-approximated by numerical simulation. Hubbell (2001) showed that neutral models that couple random demographic change with zero-sum community dynamics generate multinomial species distributions that often well approximate the observed distributions of natural communities. Zero-sum dynamics refers to the population dynamics of species in communities in which all available resources are fully exploited (Hubbell 2001).

The controversy over assembly hypotheses has continued, in part, because models based on the niche-assembly and random-assembly hypotheses can generate the same community structure [29]. Community ecologists often use species richness (number of species) and species relative abundance to characterize the structure of communities. Structure is typically described in terms of a community at a particular location, and at a particular moment in time. Used in this way, structure refers to a static, localized pattern. Ecologists have tried to discriminate among hypotheses of community assembly dynamics by generating simulated communities, based on different models, and comparing their structure [17]. Unfortunately, the structure of a community neither reveals how it was assembled, nor the mechanisms that generate the abundance distribution of species. Therefore, to determine the extent to which communities are assembled deterministically or stochastically, ecologists need information on the dynamics of species populations at community and metacommunity scales. Specifically, they need to know how species proportional abundances vary in time and space, and whether this variation can be adequately explained by random processes.

**The Effect of Niche-Differences on Variation in Ranked Abundance**

Neutral and niche-based theories of community assembly generate different predictions for how much the proportional abundance of species should vary in time and space. In the neutral theory, the proportion of a community attributed to each species is expected to be close to its proportion of the metacommunity (Hubbell 2001). Although species may alternate by chance in their rank within a community, the fraction of the community represented by each rank should be relatively stable. For example, suppose species A and B both represent eight percent of a metacommunity, and suppose they represent ranks 10 and 11 in the metacommunity. Random variation in the migration of individuals will cause a particular community to alternate in the ranks of A and B over time. But on average, they will occupy the same positions in the distribution, relative to other species. For this reason, the proportion of a neutral community represented by ranks 10 and 11 will remain fairly constant (around eight percent), despite variation in the ranked positions of A and B. Moreover, so long as species are equally able to disperse across the landscape, neutral communities of similar size and dispersal limitation should generate very similar ranked species distributions.

On the other hand, if species proportions are governed by niche differences, then the fraction of a community attributed to each rank can be quite variable in space and time. This is because the abundance of a species in niche-assembled communities depends on how well it is adapted to the local conditions, and the degree that its niche overlaps with those of other species. If niche differences are important, variation in the abiotic properties and species composition of communities can induce variation in species relative abundance above that predicted by the neutral theory. Niche-based community dynamics can also reduce variation to levels below that predicted by the neutral theory. If two geographically distant communities have the same migration probability, and experience identical or very similar abiotic conditions, deterministic interactions may regulate species relative abundance, and thus reduce variation below that expected from random dispersal.

**The Effect of Consumer Pressure on the Variation in Ranked Abundance**

One way of evaluating the importance of random processes is to experimentally manipulate environmental variables that influence non-random processes. For example, a change in the intensity of competition should affect the extent to which random factors influence species relative proportions. Processes or conditions that effectively reduce competition should bring species distributions closer to neutral model predictions, whereas factors that enhance competitive asymmetries should reduce agreement with neutral models. For example, consumer pressure has been shown to be particularly important for the strength of competition [30, 31]. Exclusion experiments, in which consumer pressure was manipulated, have revealed competitive asymmetries among prey [3, 30, 32-34]. These studies have shown that predation can generate more uniform species abundance distributions of prey and allow more species to coexist. This has implications for tests of the neutral theory. If predators alter the number and abundance of prey within communities, they may influence estimates of both metacommunity diversity and local dispersal limitation.

Here, we used long-term data on the dynamical behavior of a natural metacommunity to test the predictions of Hubbell's neutral models. We determined the temporal variation in natural species relative abundance distributions at the community and metacommunity scales, and compared this empirical variation to the variation generated by neutral models. We parameterized models using empirical data from a system of 49 small rock pools that support a diverse metacommunity of aquatic invertebrates. We also exploited natural spatial heterogeneity in predator density in the metacommunity to disentangle the relative effects of deterministic species interactions and random dispersal on community structure. While other studies have examined the importance of regional and local processes on metacommunity structure [35] Loreau and Mouquet 1999), to our knowledge, ours is the first study to analyze the dynamical relationship between individual communities and a metacommunity as a direct test of the neutral theory.

**METHODS**

**Terminology and presentation conventions**

We use the terms *natural* and *neutral* to distinguish respectively the communities and metacommunity represented by the natural system of pools from the simulated (neutral) systems generated by models. For comparative purposes we follow Hubbell (2001) in using plots of ranked relative species abundances (proportions) on the logarithm scale, where the number of ranks equals the number of species, and species are ranked in decreasing order of the logarithm of their relative proportion of a community.

**The Study System: a Natural Metacommunity**

For our analysis of natural species distributions, we used data on the invertebrate species that inhabit 49 small rock pools. The pools are located on a fossil reef, in the vicinity of the Discovery Bay Marine Laboratory, University of West Indies, Jamaica. The data were collected as part of an ongoing long-term monitoring project begun in December, 1989. Details on collection methods can be found in [36]. For a description of the physical features, water chemistry, and the environmental conditions of the pools, see [37]. Additional information and photographs of the pools are available at http://sciwebserver.science.mcmaster.ca/biology/faculty/kolasa/Research/research.html. We analyzed the contents of pool samples that were collected in December 1989, and in January of 1990, 1991, 1992, 1993, 1994, 1996, 1997, and 1998.

The 49 study pools represent a random sample of 230 pools that occur within a radius of 25 meters. The pools in the study are less than one meter apart, on average. Pool volume ranges from 0.24 to 115.00 liters (mean: 14.96, standard deviation: 21.06). Most pools contain less than 30 l. Each pool sample reasonably represents the state of the pool community at the time of collection, including species richness, the number of individuals of each species, and proportions among them.

Each sample consisted of animals contained in 0.5 l of water taken from each pool during an annual collection period. Prior to collecting a sample, we stirred the water to distribute organisms uniformly. We preserved samples in the field in 50% ethanol for species identification and counts. In the pools that contained less than 0.5 l of water, we removed a smaller volume and scaled up the specimen counts to 0.5 l. In total, the samples contain 74 invertebrate species.

**Natural Community Data Treatment**

In the neutral theory, a community is constrained to include only the individuals of trophically similar species that potentially compete for resources. The rock pool communities meet the criteria implicit in this definition. Pools are small enough to allow potential interactions between any two individuals, yet have discrete boundaries that limit interactions to organisms that occur within a pool. To meet the requirement that species represent a single trophic group, we categorized the species into trophic groups and analyzed each group separately. Therefore, for our analysis we defined a natural community to be the set of species in a particular trophic group, found in a single pool.

To determine which trophic groups are represented in the study system, and to assign each species to a specific group, we examined the trophic characteristics of families and genera as described in standard texts [38, 39]. We also used observations on the ecology and behavior of individual species (unpublished). We were able to assign all but eight of the 74 species to one of three trophic categories. The first two categories are 1) detritivores, which include dipteran larvae, nematodes, oligochaete worms, and ostracods, and 2) algal-filterers, which comprise a variety of cladocerans, ostracods, and copepods. Predators make up the third category, and include aquatic insects, such as coleopterans, tanypodid midges, and heteropterans, as well as some turbellarians, and small decapods (shrimp). All of the predators found in the pools are non-selective and are likely to feed on all prey species. In total, we identified 33 detritivores, 9 algal-filterers, 22 predators, and 8 unclassified species. We excluded unclassified species from further analysis.

Preliminary analysis showed that detritivores were, on average, the most abundant of the three trophic groups, followed by algal-filterers and predators. We limited our analysis of species distributions to detritivores to provide the most accurate estimates for community proportions and migration probabilities. To determine the effects of predator density on species richness, we analyzed both the detritivore and algal-filterer trophic groups.

**Model Parameterization**

We now describe how we estimated values for model parameters using empirical data. We defined the 49-pool natural system as a metacommunity. A metacommunity is defined as a regional set of communities linked by dispersal (Mouquet et al. 2001). Hubbell defined the species richness and proportions of a metacommunity to be a function of two parameters, the fundamental biodiversity number, or theta ($\theta$), and the maximum metacommunity size (number of individuals), $J_m$. We estimated $\theta$ and $J_m$ using a three-step process. In step 1, we calculated the average size of the natural metacommunity, and used this as a first approximation of $J_m$. We refer to this first approximation as $J_m^*$. To calculate $J_m^*$, we estimated the total number of individual organisms in each *pool* during each collection period. Note that this is different from using the total number of individuals in each *sample*. To calculate the number of individuals in a pool, we multiplied the number of individuals in a 0.5 l sample by two, and multiplied the result by the maximum number of liters in the pool. Then, for each collection period, we calculated a metacommunity size as the sum of the pool totals. Finally, we calculated $J_m^*$ as the average of the metacommunity of all the collection periods.

In step 2, we used a log-maximum likelihood equation (MLE), developed for multinomially distributed data [40], to find the best estimate of $\theta$ for a neutral metacommunity of size $J_m^*$. The MLE function compares the logarithms of the ranked values of two distributions. Here, we compared the species proportions of the natural metacommunity with those of simulated neutral metacommunities of identical size and different values of $\theta$. We used only species above the median abundance in the natural metacommunity to estimate $\theta$ as their abundance estimates are less likely to suffer from sample size effects (Hubbell 2001).

The log-likelihood of a given value of $\theta$ is:

$$l(\theta) = \log n! - \sum_{i=1}^{S} \log x_i! + \sum_{i=1}^{S} x_i \log p_i(\theta) \tag{1}$$

where $x_i$ is the abundance (number of individuals) represented by the species in rank $i$ in the natural system, $p_i(\theta)$ is the proportion of rank $i$ in a neutral metacommunity with a given $\theta$, $n$ is the total size of the natural metacommunity, and $S$ is the number of species. The result of the MLE calculation is a single value that represents how much the two distributions deviate from one another. A perfect match yields a maximum MLE value of zero. To determine the best estimate of $\theta$, we calculated a separate value of MLE for a series of neutral metacommunities, which differed in $\theta$ but not in size, until the maximum MLE value was attained.

In step 3, we calculated the maximum number of species, $S_{max}(J_m,\theta)$, generated for a neutral metacommunity of size $J_m = J_m^*$ and with $\theta$ set to the estimate calculated in step 2. We derived the equation for $S_{max}(J_m,\theta)$ from Hubbell's description of his algorithm (Hubbell 2001):

$$S_{max}(J_m,\theta) = \sum_{j=1}^{J_m} \frac{\theta}{\theta + j + 1}.$$ Using the value of $\theta$ estimated in step 2, and beginning with $J_m$ set equal to the value of $J_m^*$ estimated in step 1, we tuned the value of $J_m$ until $S_{max}(J_m,\theta)$ was equal to the number of species observed in the natural system. We repeated the above three-step process for each trophic group.

Once we had established values for $\theta$ and $J_m$, we constructed a metacommunity using the algorithm described below. From this we calculated the distribution of species proportions expected for the neutral metacommunity, which we then applied to the construction of the model (neutral) communities.

**Hubbell's Neutral Metacommunity Algorithm**

We now describe how we used the parameters estimated above to construct neutral metacommunities (Figure 1). Preliminary analyses of the mechanics of dispersal and the distribution of organisms among the pools suggested that organisms disperse widely across the metacommunity. Therefore, we modeled the study system as a spatially implicit metacommunity *sensu* Hubbell (2001). We did not pursue the alternative modeling approach, in which dispersal is limited to exchanges between neighboring communities. The proportions of species in the neutral metacommunity are determined as follows. Beginning with a single individual of a single species (metacommunity size $j = 1$, metacommunity richness $S = 1$), new individuals are added to the metacommunity until the maximum size, $J_m$, is attained. As each individual is added, it is assigned a species identity. With probability $\theta/(\theta + j - 1)$, an individual is assigned to be a new (previously unrecorded) species, where $j$ is the current number of individuals in the metacommunity. With probability $1 - [\theta/(\theta + j - 1)]$, the individual is randomly assigned to a pre-existing species. In the latter case, the probability of being assigned to a particular species, $i$, is equal to that species' current proportion of the metacommunity, $P_i$. This process is continued until all $J_m$ individuals have been added and assigned to a species.

**Hubbell's Neutral Community Algorithm**

We now describe Hubbell's neutral community algorithm, which we used to simulate neutral communities. We used Hubbell's community dynamics algorithm to simulate a separate neutral community for each pool sample. To do so, we first set the size of each simulated community equal to the number of individuals found in a respective sample, and then stochastically generated the relative proportion of each species. To generate the initial species proportions of each simulated community (colonization), we selected individuals at random from a previously established neutral metacommunity, constructed as described above. The probability of an individual in a community being assigned to a particular species, $i$, of the metacommunity is equal to its metacommunity proportion, $P_i$. The colonization process was intended to mimic the random immigration of individuals from a metacommunity to a community, with no dispersal limitation. The processes involves several steps and decisions (Figure 2A).

Once we established each neutral community, we simulated random birth, death, and migration, using a process defined by Hubbell's neutral dynamics algorithm (Figure 2B). The algorithm has four parameters: 1) community size, $J$, 2) the per capita probability of migration, $m$, 3) the disturbance level, $D$, which is the fraction (percent) of the community that is replaced each generation, and 4) the number of generations. For each neutral community, we set $J$ equal to the size of the pool sample it was intended to simulate. In our model, $J$ and $D$ do not change over time. As a result of constant $J$, the dynamics are zero-sum: before an individual can migrate into a community, or be born there, a current member of the community must first leave or die. In addition to the above parameters, and the initial distribution of community species proportions, the algorithm requires a pre-established metacommunity species distribution, to provide the $P_i$ of each species.

The value of $D$ influences the number of species found in neutral communities. The probability that rare species are extirpated in a community is proportional to $D$. At very high values of $D$, only the most abundant species in the metacommunity are likely to be present within a community. We estimated the natural level of disturbance that occurs in the pools over the course of a year by calculating the average decline in the size of the pool samples. We only considered those instances in which the sample for a pool was smaller than the previous sample. When the pool communities decreased in size, the reduction was 60.45% on average from one annual sample to the next. The per generation disturbance level should be less than the annual level, because the pool organisms undergo multiple generations per year. Generation times for the organisms in the pools vary considerably among species, ranging on the order of days for some species to months for others. Our goal was to reasonably approximate the average natural level of disturbance in the model communities. Therefore, we conservatively estimated the value of $D$ for the natural system at 10% per generation.

To simulate the processes of birth, death, and immigration, we randomly changed the species identity of community members to species chosen from either the community or metacommunity. In each generation, $D \times J$ individuals are randomly chosen from the community for replacement. With probability $m$, an individual is replaced by an individual of a species drawn at random from the metacommunity. With probability $1-m$, it is replaced by an individual from a species drawn at random from within the community. If the replacement comes from the metacommunity, the probability that a particular species, $i$, will be selected is equal to its metacommunity proportion, $P_i$. If the replacement comes from the community, the probability that a particular species will be selected is equal to its proportion of the community. A generation ends when the identity of all of the $D \times J$ individuals has been reassigned.

The replacement process is continued until a dynamic equilibrium in species community proportions is reached. The dynamic equilibrium is achieved when the distribution of ranked proportions becomes stable such that each rank varies in its metacommunity proportion by less than 10% each generation. Once the dynamic equilibrium is attained, we record the species proportions every 1,000 generations until 1,000 separate species distributions have been recorded for each neutral community.

**Estimating Migration Probability**

In this section, we describe how we derived estimates of the probability of migration for the pools in our study. Neutral theory predicts that the number of species in a community, and the proportion of the community represented by each species, depends on the community size ($J$, number of individuals), the number of species in the metacommunity, $S$, the proportions of each species in the metacommunity, $P_i$, and the probability of migration, $m$, between community and metacommunity. We set the value of $J$ to the size of each sample, and established values for $S$ and $P_i$ as described above for metacommunity construction. Different values of $m$ result in different values of species richness and proportional abundance. To find the value of $m$ that best fit the species distribution of each pool, we calculated the predicted species richness and proportions of neutral communities, and compared them statistically to those derived from the pool samples. Each pool sample contains a specific number of individuals (community size) and species (richness), providing a size-richness distribution for each pool (one value of richness for each sample). We compared the distribution of species richness in the samples to those of neutral communities of the same size using a Wilcoxson matched pairs test. Each pool sample and neutral community also generates a distribution of ranked species proportions. For each sample, we compared the average proportion of each rank in the samples to those of neutral communities of identical size, using a Chi-square test. For both richness and proportions, we used four different values of *m* to generate the predicted values of neutral communities: 1.0, 0.10, 0.01, 0.001.

To find the estimate of *m* that best characterizes each pool, we considered the species richness and proportions together. For each pool, the best estimate of *m* is the value that generates the species richness and distribution of proportions in neutral communities that are not significantly different from those of the pool's samples. Thus, for each value of *m*, we compared the results of the Wilcoxson matched pairs test and Chi-square test, and chose the value of *m* that produced non-significant statistical results for both tests. If more than one value of *m* produced this result, we estimated the migration probability of a pool as the range bounded by the low and high values of *m* thus attained. If a value of *m* produced a significant result for one of the tests, but not for the other test, then we estimated *m* based on a single test.

**Community Dynamics: Coefficient of Variation in Rank Proportions**

To test how well neutral models reproduce the dynamics of natural communities, we examined the variation in species proportions within each pool. To measure this variation, we used the CV in the fraction of the community represented by each rank, expressed as a percent ($(SD/mean) \times 100$, where SD is the standard deviation). To compare the CV in species proportions between natural and neutral communities, we used the neutral community distributions that represent the previously established estimates of *m* for each pool.

**Effect of Hydrological Conditions on Migration and Population Dynamics**

The probability of migration, and extent to which the population dynamics are zero-sum, are likely to have been affected by natural variation in rainfall. The pools lost water to evaporation, and were refilled during storms by rain and sea water depending on their position, which changed their physical and chemical properties such as thermal regime, nutrient concentrations, and oxygen concentration (Therriault and Kolasa 1999). Substantial movement of organisms among the pools occurred as they were carried in water that overflowed from the pools during rain (D. Jenkins, *pers. comm.*). In years that received less rain, the rate of migration among pools may have been reduced. Stochastic variation in water volume and chemical properties also changed the carrying capacity of the pools, potentially influencing community dynamics.

We estimated the relative magnitude of variation in hydrological conditions by comparing the number of pools that were dry during each collection period. Most of the time, fewer than 3 pools were dry during a collection period. However, 9 pools were dry in 1993, and 28 pools (57%) were dry in 1994. The average volume of all the pools was likely reduced during these two years. To estimate the degree to which variation in pool volume influenced the migration probability and community dynamics of the pools, we compared the coefficient of variation (CV) of each rank for the selected pools to a subset of the data that did not include data from the low-rainfall years of 1993 and 1994.

**Effect of Predators on Prey Species Distributions**

As a further test of the importance of niche-differences for species distributions, we analyzed the effect of predators on prey communities. High densities of predators may alter the richness of prey communities, such as by affecting the intensity or outcome of competition [30, 41]. If so, predators could influence the importance of random processes for community distributions.

**Estimating Predator Density**

Predators varied in density among the pools from year to year. In total, they are present in 52% of the samples. When present in a sample, predators average about 10% of the total number of individuals. To assess the impact of predators on prey species, we first needed to calculate the probability of not detecting predators in a pool when they were present. The probability of observing $c$ or fewer individuals of a species in a sample of size $n$ taken from a community of size $N$, is

$$P(x \leq c) = \sum_{x=0}^{c} \frac{\binom{a_i}{x}\binom{N-a_i}{n-x}}{\binom{N}{n}} \qquad (2)$$

where $a_i$ is the abundance of species $i$ in a community and $x$ is the observed number of individuals of species $i$ [17]. For a given $a_i$, the probability $P(1)$ of observing at least one individual of species $i$ (or any predator, in our case) is $1-P$, with $c$ set to zero in equation (2). In this case, equation (2) reduces to:

$$P(1) = 1 - \frac{\binom{N-a_i}{n}}{\binom{N}{n}} \qquad (3)$$

The probability of detecting a predator is inversely related to sample size; larger samples can detect predators at lower predator densities. In principle, sampling fails to capture predators that occur below some minimum density. To discover what that minimum density was, we asked the question: if the fraction of a community represented by predators is $F$, what is the minimum value of $F$ needed for there to be a 95% chance of capturing at least one predator in a sample ($P(1) = 0.95$)? We denote the minimum value of $F$ as $F^*$.

To answer this question, we considered the worst case, of a sample containing a small number of individuals. The number of organisms in samples that lacked predators range from a single individual to a maximum of 25,317 individuals (mean 846.4). We estimated the value of $F^*$ for one of the smallest samples, which contained 11 individuals. We set $N$ in equation (3) to $2nV$, where $V$ is the volume of the pool in liters, and set $n$ to the number of individuals in the sample. Then, by varying $a_i$ until $P(1)$ is at least 0.95, we were able to estimate the value of $F^*$. For a sample size of 11, the value of $F^*$ needed to generate a $P(1) \geq 0.95$ is 21.5%. The sample we used to estimate $F^*$ is much smaller than most of those that lack predators: 92.6% of the samples that lack predators are larger than 11 individuals. We defined samples that lack predators as low-predator-density (LPD) samples, representing communities in which predators account for $\leq 21.5\%$ of the total number of individuals. High-predator-density (HPD) samples are defined as containing $> 21.5\%$ predators.

**Effect of Predator Density on Estimates of Migration Probability**

To test the effect of predator density on estimates of migration probability, $m$, we compared the species richness and species proportional abundance of the LPD and HPD samples taken from each pool. We followed the same approach described above in the section on migration probability. We compared the average species proportions of the neutral communities to the average species proportions derived from the LPD and HPD samples of each pool, using a Chi-square test. To test for an effect of predators on richness, we compared the richness of LPD and HPD samples of each pool using a Wilcoxson matched pairs test.

Although the samples were drawn from a random sample of 49 pools, it's possible that, by coincidence, the structure of prey communities in the HPD samples simply reflects variation in detritus or abiotic conditions. If so, prey community size in HPD samples should be larger than in LPD samples, and the effect of predators on prey community structure would be confounded by other factors. On the other hand, if competition among prey species is reduced by predators, as a result of reduced prey population sizes, then HPD samples should have lower prey densities, on average. To determine if HPD samples are biased toward large community sizes, we compared the average size of the detritivore communities in LPD and HPD samples, using a Wilcoxson matched pairs test.

**RESULTS**

We chose three pools to illustrate the results of our analysis. Qualitatively similar results hold for the other 46 pools. To show results for a range of pool volumes, we chose pools that represent three pool volume classes: < 10 l, 10 to 20 l, and > 20 l. Pool #12 is small, with a maximum volume of 3.0 l and an average sample size of 82.9 individual organisms (min. 9, max. 263). Pool #9 is characteristic of large pools, with a maximum volume of 29.0 l, and an average sample size of 407.9 (min. 2, max. 1,529). Pool #29 falls between the two other pools in volume and size, with a volume of 20.0 l, and average size of 236.4 (min. 20, max. 307; Figure 3)**Community Species Distributions and Migration Probabilities**

The distribution of species proportions closely approximate an exponential distribution, in particular for pools #9 and #12 (Figure 4). Although individual species ranks of natural communities often agreed with their respective ranks for a specific neutral community, no single neutral distribution fit the entire distribution curves of the natural communities. For example, most of the different species ranks in pools #9 and #12 varied in their agreement with curves for migration probabilities of either 1.0 and 0.10. The average proportion of the last rank of pool #9 fell between curves for $m = 0.01$ and 0.001. All of the neutral community species distributions were significantly different from the natural distributions for migration probabilities greater than 0.0 (Table 1). Also, species richness (number of ranks) in the pools tended to be lower than predicted by the neutral theory (Figure 4, Table 1). The margins of error for individual points was high, and in most cases, the curves representing neutral communities with migration probabilities between 1.0 and 0.01 could not be distinguished statistically. These results illustrate that the pools are not well characterized by a single migration probability, and suggest that other factors besides dispersal are important to community structure. Based on the comparison of natural and neutral species distributions and species richness (Table 1), pools #12 and #29 were best approximated by curves representing a migration probability of 0.01, while pool #9 (the largest pool of the three) was best approximated by a migration probability of 0.10.

**Metacommunity Species Distributions**

The value of θ estimated by the maximum likelihood method for the detritivore metacommunity is 1.9. The corresponding neutral metacommunity distribution provides a good fit to the empirical pattern ($R^2 = 0.9528$), and the curves for the natural and neutral distributions are statistically indistinguishable (Figure 5). In particular, the ranked proportions of common detritivores are generally close to their predicted values. However, the neutral distribution did not fit the natural distributions of rare species. Over a third of the species with low metacommunity proportions (i.e. < 0.01%) were consistently more abundant than predicted (Figure 5).

**Community Dynamics: Variation in Rank Proportions**

The variation in rank proportions was quite different in the natural pools than in neutral communities of identical size (Figure 6). In the neutral communities, the CV of community proportion of all but the rarest species tends to increase linearly in proportion to rank. But in many natural communities (such as pools #12 and #29 in Figure 6), the most abundant species are often less variable in their community proportion than predicted by neutral models, while rare species are much more variable. Most of the CV values for pool #12 are bracketed by those of curves that represent the two values of *m* that best characterized the distribution of species in the pool's natural community. This illustrates the inability of a single migration probability to characterize the natural distributions.

**Effect of Hydrological Conditions**

Variation in pool volume influenced the variation in species rank proportion. We found that when the driest years were not included, the CV in species proportions of all but the top ranks was lower compared to their proportions in the full data set (Table 2). However, in most cases, the effect of the dry years was not enough to substantially improve the fit of natural and neutral CV values. Thus, while variation in migration probability and community dynamics probably contributed to the observed variation in species proportions, it was not sufficient to explain the observed patterns.

**Predator-Prey Interactions**

The influence of predators on prey richness differed for the two prey groups. High predator densities (HPD) significantly increased mean species richness for detritivores from 2.26 to 4.25 (Wilcoxon matched pairs test, $p < 0.00005$), but the effect on algal-filterers was not significant (from 1.26 to 1.93; $p = 0.4631$). We found no statistical difference in the sizes (combined abundance) of detritivore communities between low predator density (LPD) and HPD samples (Wilcoxson matched pairs test, $p = 0.1630$). Thus, we could not find evidence that differences in richness between LPD and HPD samples were caused by a difference in the sizes of LPD and HPD samples. There was a tendency for detritivore densities in HPD samples to be lower than in LPD samples: the average community size was 430.4 in HPD samples and 558.7 in LPD samples. This suggests that the increased species richness of detritivore communities is not related to high prey densities.

Predator density influenced all aspects of detritivore community structure. In addition to increased species richness, the proportions of detritivore species in HPD samples were more evenly distributed (Figure 7). While the magnitude of the effect of high predator density varied among the 49 pools, LPD samples most often contained fewer detritivore species, and the proportion of of the community represented by the lower ranks (less common species) was significantly less than in HPD samples (Chi-square $p < 0.00001$ for each of the 3 pools shown in Figure 7). Predator density also influenced the composition of the detritivore community samples. Six detritivore species were found only in HPD samples, while two species occur only in LPD samples. In each case, the organisms limited to HPD or LPD samples represent very rare species, with average metacommunity proportions between $7 \times 10^{-6}$ and $7 \times 10^{-5}$.

Differences in predator density also affected the estimate of θ for the detritivore metacommunity. The value of θ estimated for the entire system (all samples together) was 1.9. When only HPD samples are considered, θ = 2.6, reflecting the higher average richness and species proportions. When only LPD samples are considered, θ = 1.28. While the statistical error around the values in Figure 7 is substantial, given the consistency of the above patterns, the effects of predator density on species distributions can not be completely explained by random sampling error. We concluded that predator density exerts a substantial influence on detritivore community structure.

**DISCUSSION**

We tested the ability of Hubbell's zero-sum neutral models to predict the distribution of species proportions in a natural metacommunity, and in individual communities. Our results did not agree with the predictions of the neutral theory at both the community and metacommunity scales. At the community scale, the neutral theory proposes that variation in species proportions and species richness arises through variation in community size and dispersal limitation. Because the distribution of species proportions is much more sensitive to migration probability than it is to community size (Hubbell 2001), the neutral theory predicts that migration probability largely determines the agreement between the proportions of species in a community and their metacommunity proportions. Specifically, the neutral theory assumes that all species in a metacommunity have an equal probability of dispersing to a community, and therefore the distribution of species proportions is controlled by a single migration probability. If true, then neutral models should be able to generate species distributions that fit natural distributions, once the natural migration probability has been determined. At the metacommunity scale, species proportions are assumed to be largely immune to variation among communities in migration probability.

In our study, natural community species distributions did not conform to those generated by neutral models for any single migration probability (Figure 4). The variation (CV) in the fraction of a community represented by each species rank often deviated strongly from the variation predicted by neutral models (Figure 6). Differences between natural and neutral communities in the abundance distribution of species were also apparent at the metacommunity scale. Rare species (38% of the natural metacommunity) were more abundant than predicted by the neutral theory (Figure 5). Several factors, including variation among pools in migration probability and community dynamics, variation among species in tolerance for predation, variation among species in their tolerance to abiotic conditions, and differences in competitive ability, may explain the discrepancy between the models and observed patterns.

The variation in the fit of common and rare species to neutral species distributions reflects differences in their probability of survival in different pools. In particular, predator density influenced the ability of different species to coexist. Pools that contained high predator densities (HPD) had both higher prey species richness and more evenly distributed prey species proportions (Figure 7). These patterns suggests that abundant species were more affected by predation and, perhaps, that high predator density reduced competitive asymmetry among prey. If true, predators may increase the importance of random processes for prey community structure by reducing the influence of species niche differences. The improved fit of predicted neutral community proportions with those of the natural communities in HPD samples supports this idea.

Annual variation in rainfall appears to be partly responsible for the high variation (CV) in community proportions observed for rare species in the natural communities. Variation in the number of dry pools encountered during sample collections influenced the CV in community proportions (Figure 7, Table 2). This result could be attributed to variation in the migration probability of the pools, as a result of reduced water flow between pools. Or it could be related to an increase in the isolation of some pools in years of low rainfall when the fraction of pools that are dry increases. Increased isolation could occur if the more numerous small pools facilitate migration by acting as stepping stones between large pools. On the other hand, the observed variation in the community proportions within pools could be attributed to deviation from zero-sum population dynamics, as a result of fluctuations in pool carrying capacity. For example, immediately following a rain, pools contain a greater abundances of resources needed for population growth. If population sizes are frequently below their carrying capacity, then variation in the intrinsic growth rate among species will cause the dynamics of population growth to be unconstrained by the zero-sum condition. For example, rare species have an advantage when carrying capacity is unsaturated, because they experience lower levels of intraspecific competition than do species with large populations.

Previous work has demonstrated that species in our study system differ in their tolerance to variation in abiotic conditions [37, 42]. Here, we showed that species relative abundance in the pools is also influenced by predator density. It appears that variation in predator density and abiotic factors interacts with random factors, such as dispersal, to generate variation within and among pools in the average proportion of each abundance rank. Thus in our study system, variation in the distribution of community proportions is not attributable solely to random processes, but it is likely to arise through complex interactions between random and deterministic processes. The relative importance of any one factor depends on the physical and biological conditions found within individual pools.

At the metacommunity scale, Hubbell's neutral model provided a reasonably good approximation of the distribution of common species in the pools. However, rare species (38% of the metacommunity) were consistently more abundant than predicted. This result contrasts with every one of the 10 examples of natural metacommunity distributions in Hubbell's monograph, in which rare species are typically under-represented, and seldom more abundant than predicted by neutral models. Our results also conflict with Hubbell's theoretical prediction for the abundance of rare species in communities and on islands. Hubbell theorized that dispersal limitation, and the vulnerability of small populations to extinction, combine to cause rare species to be under-represented on islands and within communities, relative to their metacommunity proportions (Hubbell 2001; chapter 5). Hubbell depicted the metacommunity dominance diversity curves for a wide range of organisms, including fish, birds, bats, and bees, that fit the predicted pattern (Hubbell 2001; chapter 9), in which rare species are under-represented in natural metacommunities. Why then, are rare species in our study *over-represented* in the metacommunity?

We suggest that the same factors that influence species distribution at the community scale may also explain the abundance of rare species in the metacommunity. Ecological theory and empirical studies have shown that environmental heterogeneity exerts a strong influence on species diversity in many kinds of communities [2, 43-48], including small pond invertebrates [6, 35]. Spatial variation across the landscape in the biotic and abiotic conditions of the pools can enhance opportunities for less competitive species, increasing their abundance at the metacommunity scale over that expected by neutral theory. In particular, predator density was important to prey diversity in our study. Recall that 6 species were found only in high predator density samples. High predator density improved the survival of rare species in a significant number of pools, elevating their representation in the metacommunity. This result underscores the potential of cross-trophic interactions to influence community structure and the proportional abundance of species in the metacommunity [49].

To be fair, we note that Hubbell does acknowledge that differences among species can affect species distributions within communities [20, 50]. However, in the tropical forests he studies, Hubbell suggests that the effect of species differences is highly localized, and therefore unlikely to influence metacommunity species patterns [51]. By contrast, our results suggest that individual responses to environmental heterogeneity can have a pronounced effect on species metacommunity proportional abundance.

Our results suggest that the community assembly processes offered by the neutral theory are insufficient to account for the assembly process in the natural system we studied. For example, the greater than expected variability of rare species is compatible with predictions of hierarchy theory (Waltho and Kolasa 1994, Kolasa and Li 2003). It appears that the primary reason for this inadequacy is that the probability of survival is not constant as required by the neutral theory, but depends on species characteristics and the biotic and abiotic conditions of each pool. Differences in survival influence estimates of the migration probability of the pools. Thus, we suggest that models might provide a better fit to the natural species distributions of our system by allowing the probability of survival to vary among species. However, our analysis of predator effects suggest that the procedure necessary for assigning the survival probability of species may be more complex than a simple random assignment. The occurrence of species in the pools is far from random, and depends on both abiotic conditions and the density of predators. Therefore, to improve predictions of the distribution of species in the natural communities we studied, it may be necessary to use a spatially explicit metacommunity model which includes a parameter for habitat conditions that influences survival and varies through time and space. There are many ways to incorporate the effect of habitat heterogeneity into models [48, 52]. Introducing variation in survival at the species level would allow models to consider both neutral and deterministic dynamics.

As a null model for community dynamics, the neutral theory provides predictions that can help ecologists uncover the relative importance of random and non-random factors for community structure. However, data that captures the dynamical behavior of ecological systems are required to distinguish the effects of different factors. In addition, it's important to recognize the influence of environmental heterogeneity on estimates of migration and biodiversity. Used judiciously, neutral models can reveal the influence of deterministic factors that might otherwise be missed by other approaches.

Table 1. Estimated pool migration rates (*m*) for three pools. Statistical probability values (*P*) represent the results of comparisons between natural and neutral (simulated) communities in ranked abundance (Chi-square test; CS) and species richness (Wilcoxson matched pairs test; WCX) for four different migration probabilities. Each pool represents a different pool size class, based on volume. See text for details.

Pool 9

| Neutral $m$ | Species Proportion $P$ (CS) | Richness $P$ (WCX) | Estimated Pool $m$ |
|---|---|---|---|
| 1.0 | p < 0.0000 | 0.0299 | |
| 0.10 | p < 0.0000 | 0.1508 | 0.10 |
| 0.01 | p < 0.0000 | 1.0000 | |
| 0.001 | p < 0.0000 | 0.1763 | |

Pool 12

| Neutral $m$ | Species Proportion $P$ (CS) | Richness $P$ (WCX) | Estimated Pool $m$ |
|---|---|---|---|
| 1.0 | p < 0.0005 | 0.0179 | |
| 0.10 | p < 0.0023 | 0.1763 | |
| 0.01 | p < 0.5003 | 0.6121 | 0.01 |
| 0.001 | p < 0.0001 | 0.1834 | |

Pool 29

| Neutral $m$ | Species Proportion $P$ (CS) | Richness $P$ (WCX) | Estimated Pool $m$ |
|---|---|---|---|
| 1.0 | p < 0.0000 | 0.0179 | |
| 0.10 | p < 0.0001 | 0.0209 | |
| 0.01 | p < 0.1421 | 0.1234 | 0.01 |
| 0.001 | p < 0.0000 | 0.7531 | |

CS = Chi-square test
WCX = Wilcoxson matched pairs test

Table 2. Effect of hydrological conditions on the variation in community proportions. The greatest number of dry pools occurred during 1993 (9) and 1994 (28), which represent the years of lowest rainfall. When data for these years are excluded, the value of CV in the fraction of the community represented by each rank is reduced. The last column shows the percent change in CV between columns 2 and 3.

Pool #9

| Rank | All Years CV | CV sans 1993 & 1994 | Change in CV |
|---|---|---|---|
| 1 | 30.96 | 33.35 | 7.71% |
| 2 | 54.70 | 58.92 | 7.72% |
| 3 | 91.35 | 88.02 | -3.65% |
| 4 | 131.59 | 119.94 | -8.86% |
| 5 | 220.39 | 207.23 | -5.97% |
| 6 | 235.18 | 218.97 | -6.89% |
| 7 | 282.84 | 264.58 | -6.46% |

Pool #12

| Rank | All Data CV | CV sans 1993 & 1994 | Change in CV |
|---|---|---|---|
| 1 | 36.24 | 36.77 | 1.47% |
| 2 | 78.05 | 66.98 | -14.18% |
| 3 | 93.65 | 79.02 | -15.62% |
| 4 | 93.83 | 66.58 | -29.05% |
| 5 | 148.66 | 120.02 | -19.27% |
| 6 | 185.82 | 155.54 | -16.30% |
| 7 | 203.68 | 172.25 | -15.43% |
| 8 | 282.84 | 244.95 | -13.40% |
| 9 | 282.84 | 244.95 | -13.40% |

Pool #29

| Rank | All Data CV | CV sans 1993 & 1994 | Change in CV |
|---|---|---|---|
| 1 | 33.54 | 36.47 | 8.75% |
| 2 | 68.20 | 71.61 | 5.00% |
| 3 | 74.65 | 74.52 | -0.17% |
| 4 | 134.33 | 151.01 | 12.42% |
| 5 | 282.84 | 264.58 | -6.46% |
| 6 | 282.84 | 264.58 | -6.46% |
| 7 | 282.84 | 264.58 | -6.46% |

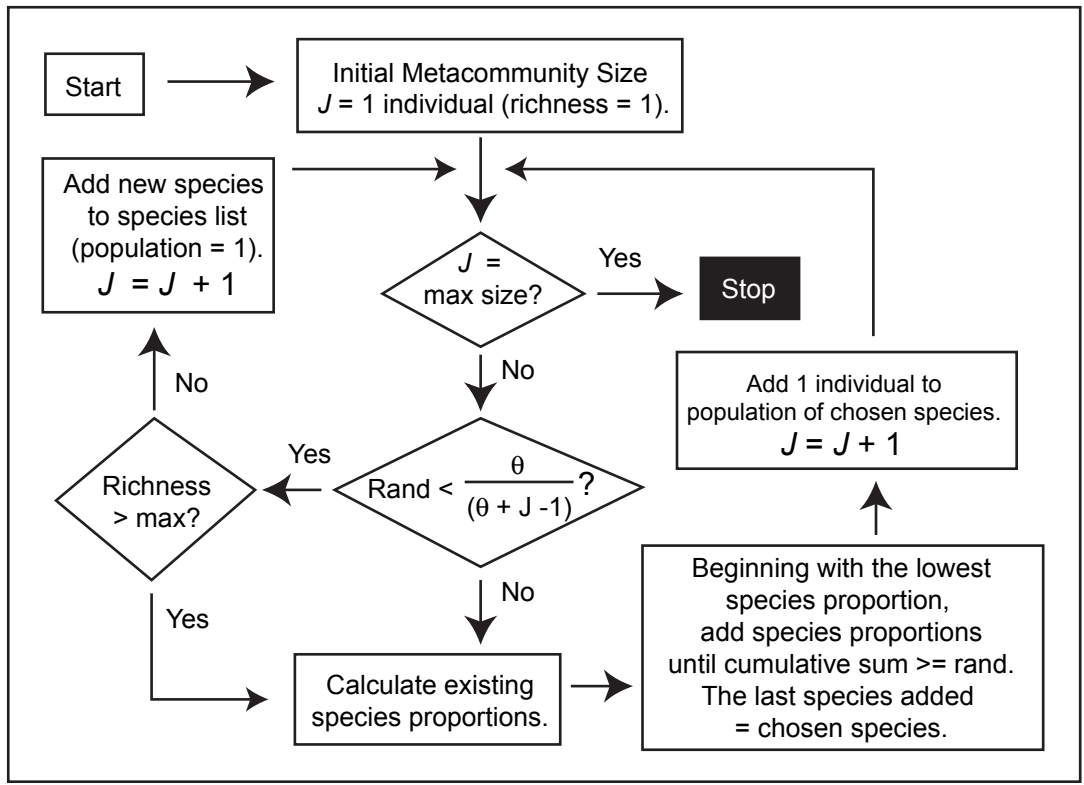

Figure 1. Metacommunity Generation Algorithm

The richness and proportions of species in a neutral metacommunity are determined by stochastically choosing whether each individual represents a new or pre-existing species. The final values of richness and proportions are a function of q and the metacommunity size, Jm. "Rand" = random number between 0 and 1. See text for details.

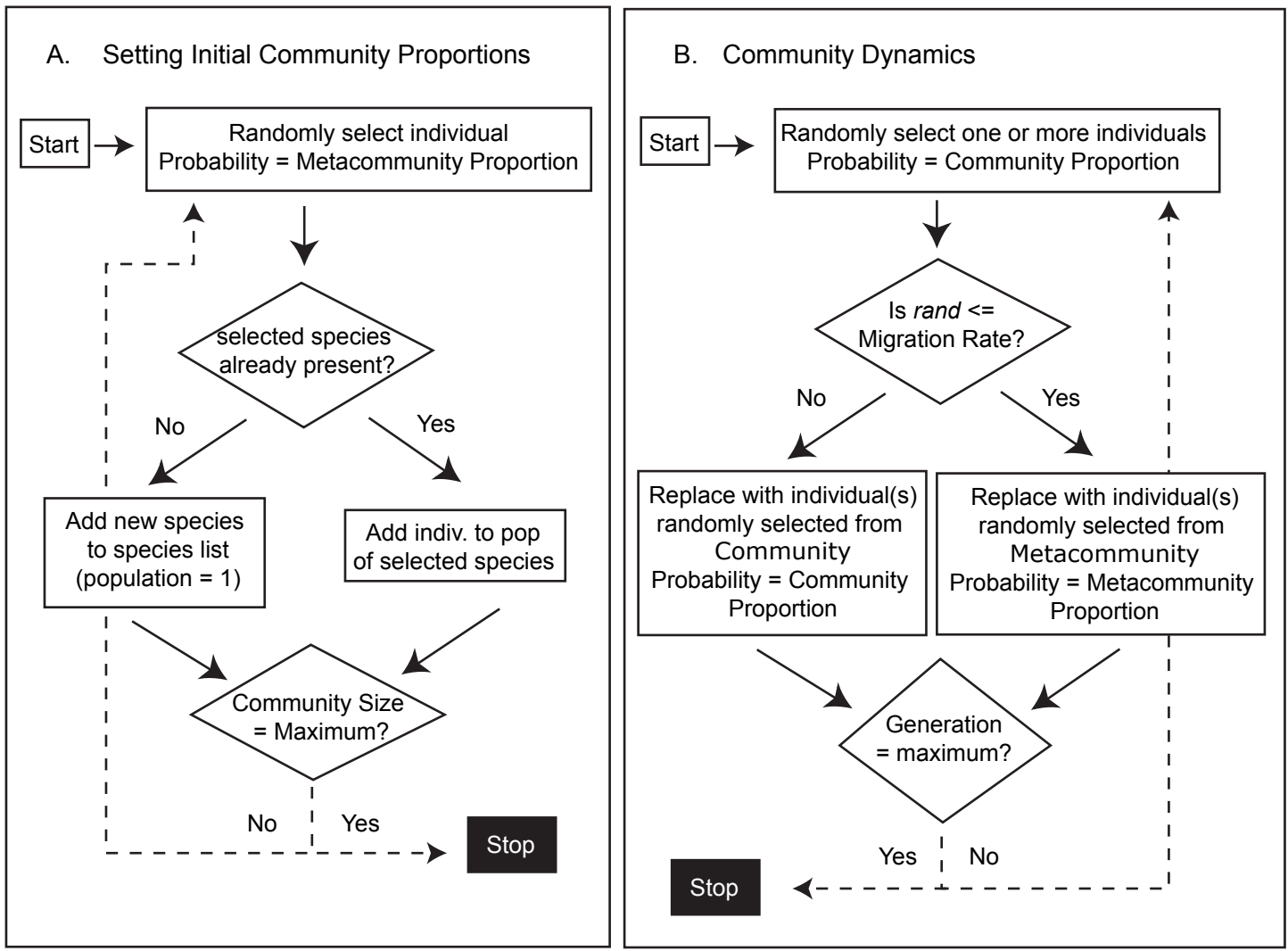

Figure 2. Graphical Depiction of Neutral Model Dynamics

2A: Initial community proportions (colonization) are determined by randomly drawing from the Metacommunity. For each species in the metacommunity, the probability, p, of being selected for the community is equal to its metacommunity proportion. 2B: Community proportions change each generation. Individuals of a species "die" with probability equal to a species community proportion. These are replaced by individuals selected randomly from either the metacommunity (p = migration rate, m), or the community (p = 1 - m). "rand" = random number between 0 and 1.

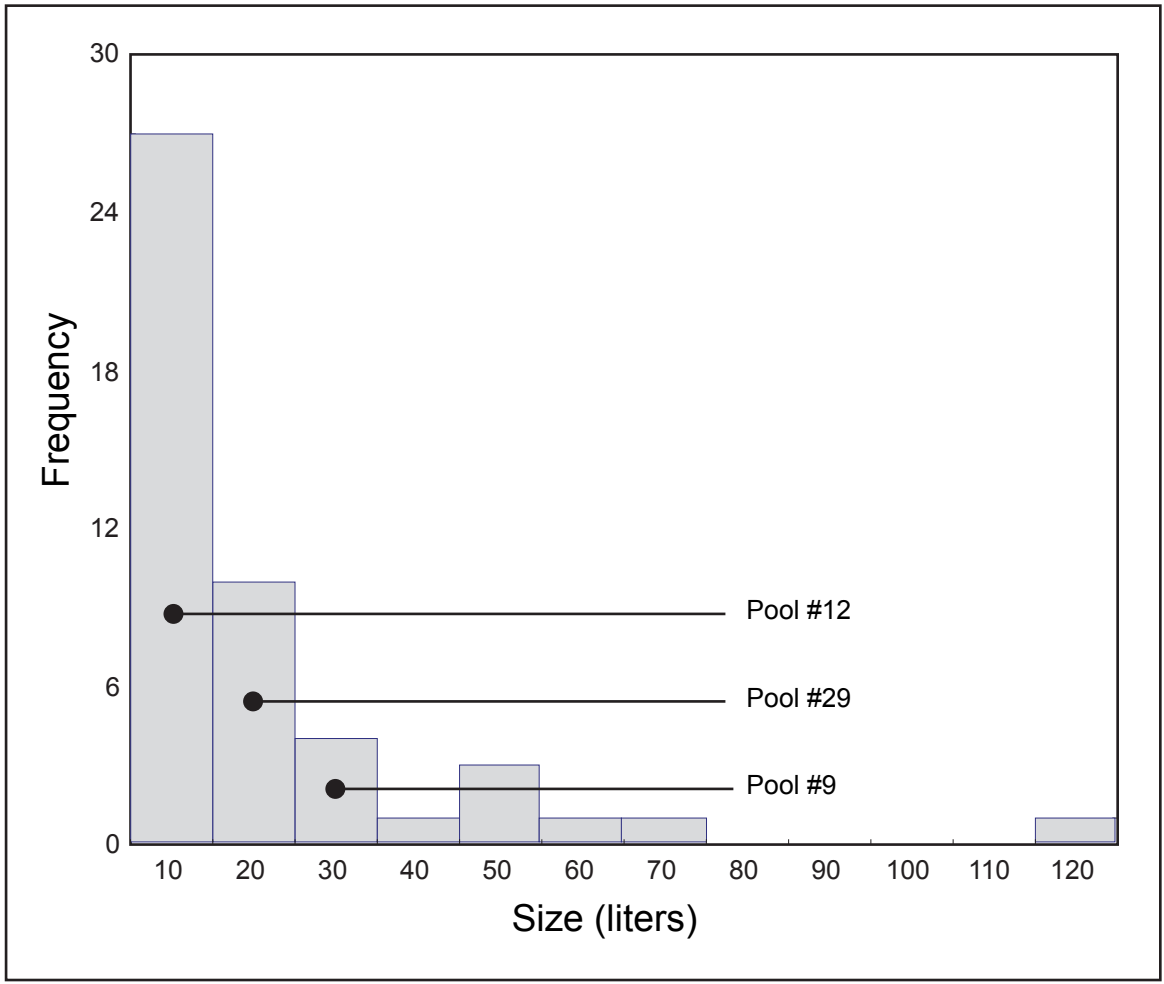

Figure 3. Distribution of Pool Volume

Each bar shows the number of pools in each volume class. Filled circles indicate the position of the three pools shown in the other figures, relative to the overall distribution.

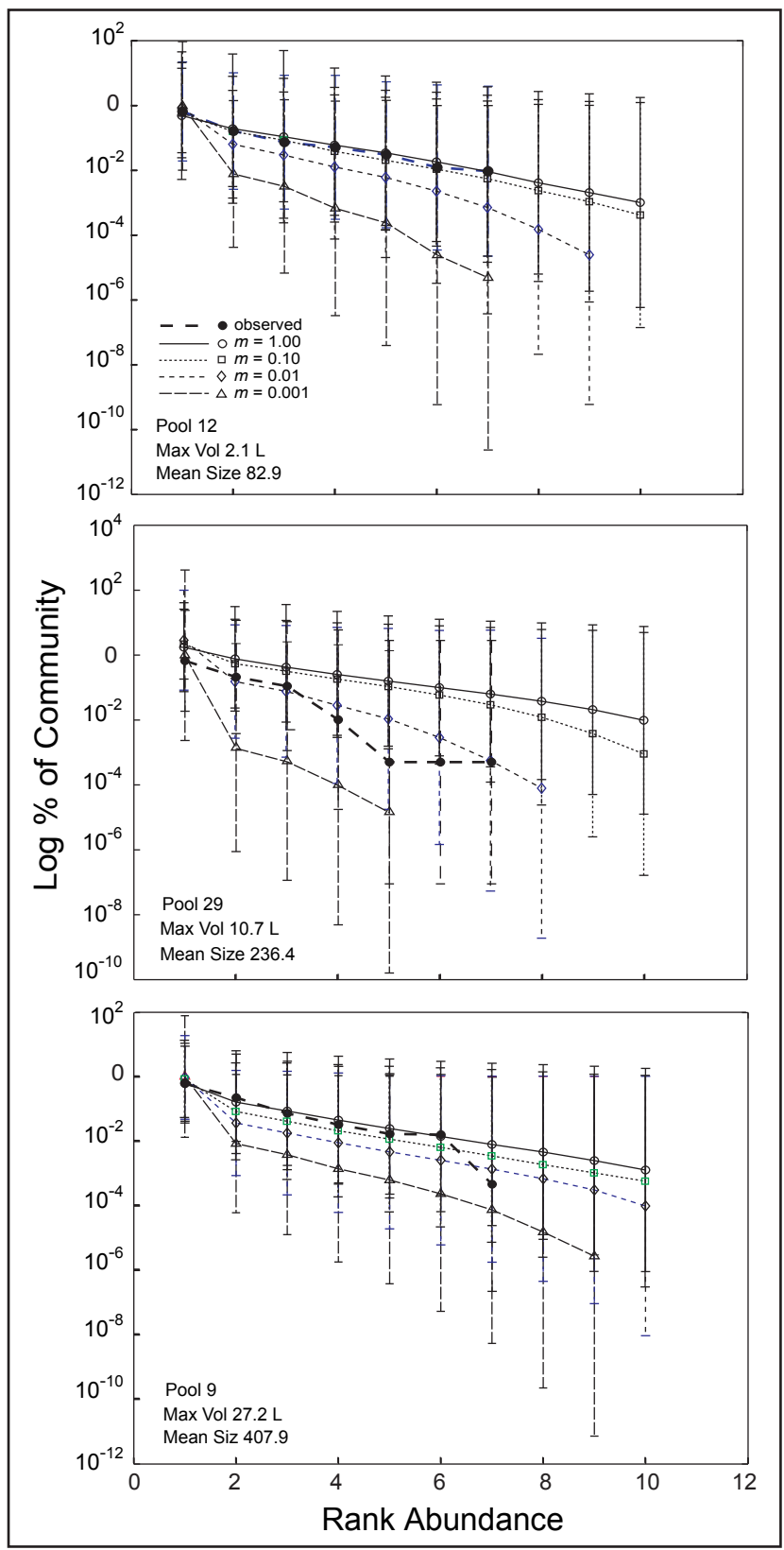

Figure 4. Community Proportions and Probability of Migration

The logarithmically transformed ranked distributions of species average community proportions is shown for each of three pools, representing different pool size classes (filled circles, dashed line). Open symbols with solid lines represent the average proportions of neutral communities, which are identical in size to the respective pools. Each neutral community represents the species proportions predicted for different values of the migration probability, m. Open circles: m = 1.00; open squares: m = 0.10; open diamonds: m= 0.01; open triangles: m = 0.001. Error bars represent standard errors.

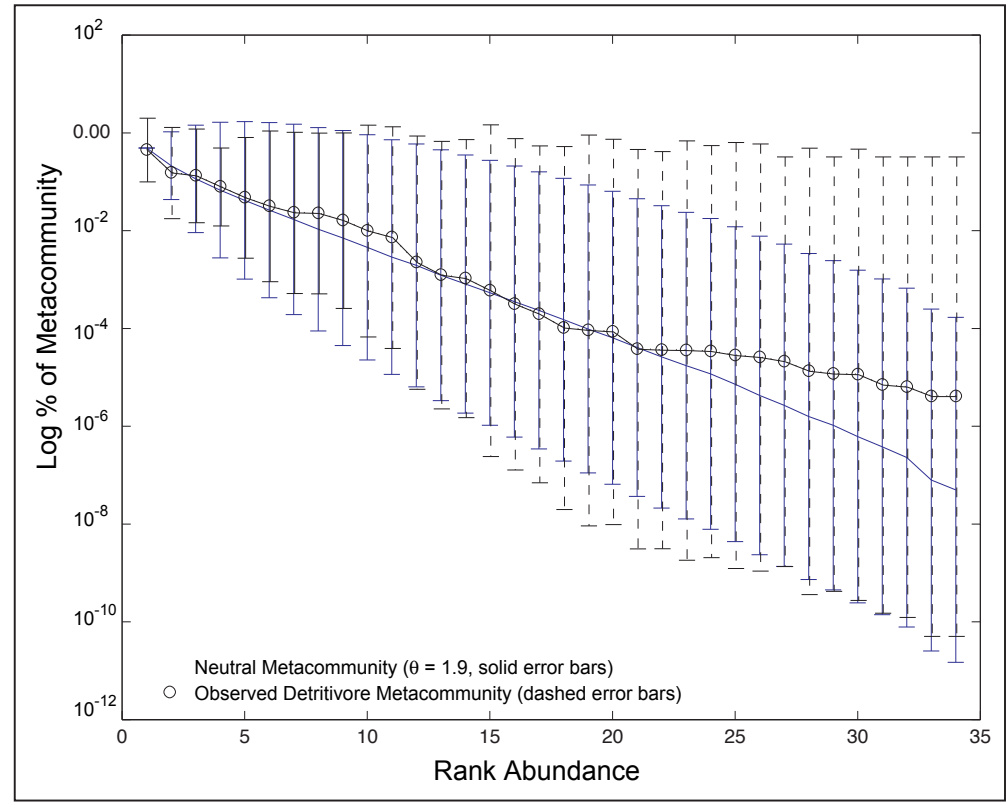

Figure 5. Metacommunity Proportions

The average metacommunity proportions are shown for ranked species in a neutral metacommunity (solid line) and a natural metacommunity of 49 pools (open circles). For the neutral metacommunity, q = 1.9 and Jm = 34 million. Neutral metacommunity values represent the mean and one standard deviation of 1,000 simulations (solid error bars). Natural metacommunity values represent the mean and one standard deviation of 9 collection periods (dashed error bars).

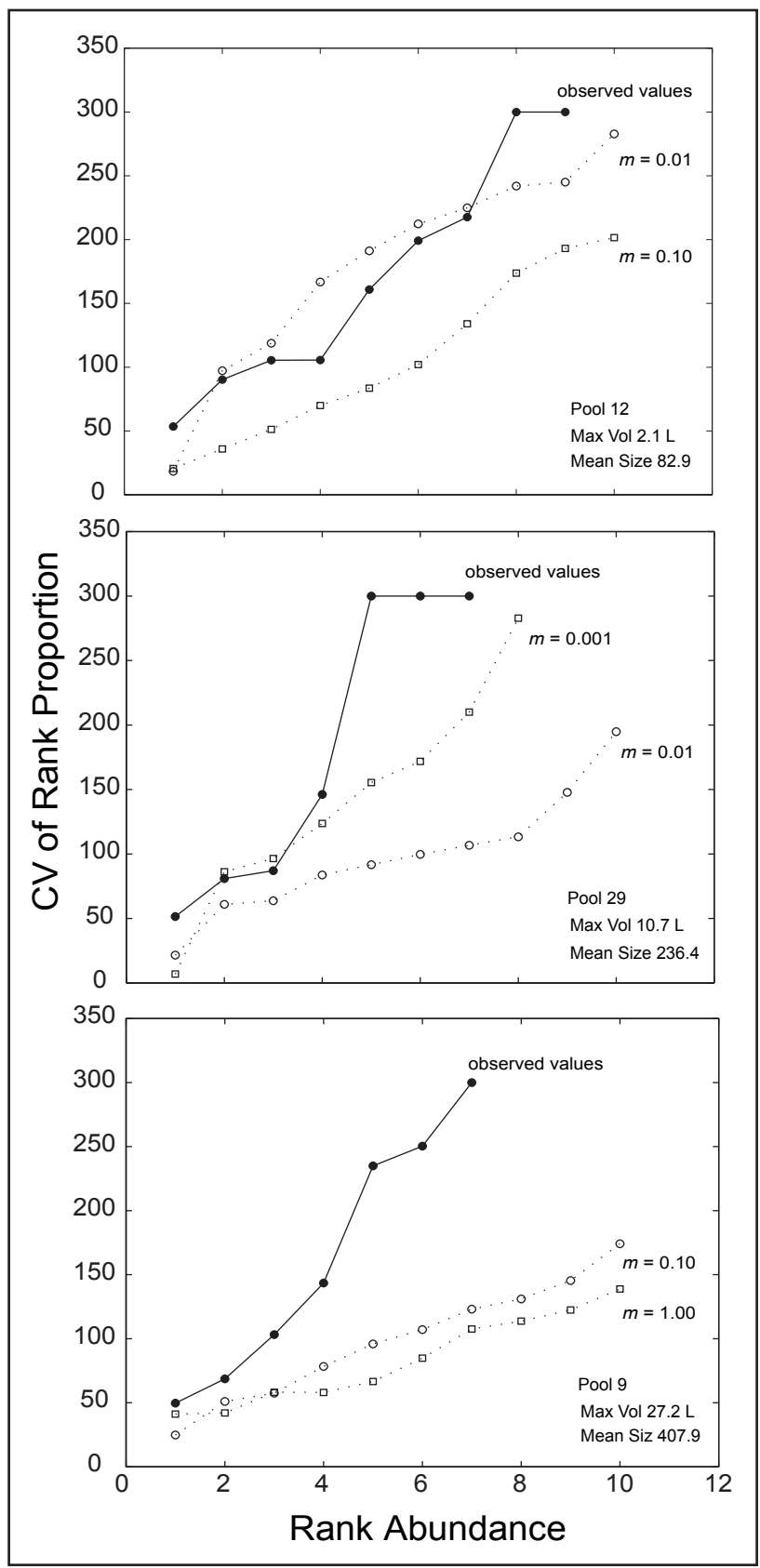

Figure 6. Coefficient of Variation in Species Proportions

The ranked distributions of the coefficient of variation (percent CV) in community proportions is shown for each of three pools, representing different pool size classes (filled circles). Open symbols represent the CV of neutral communities, which are identical in size to the respective pools. Each neutral community represents the CV predicted for different values of migration probability, m, as indicated in the figure.

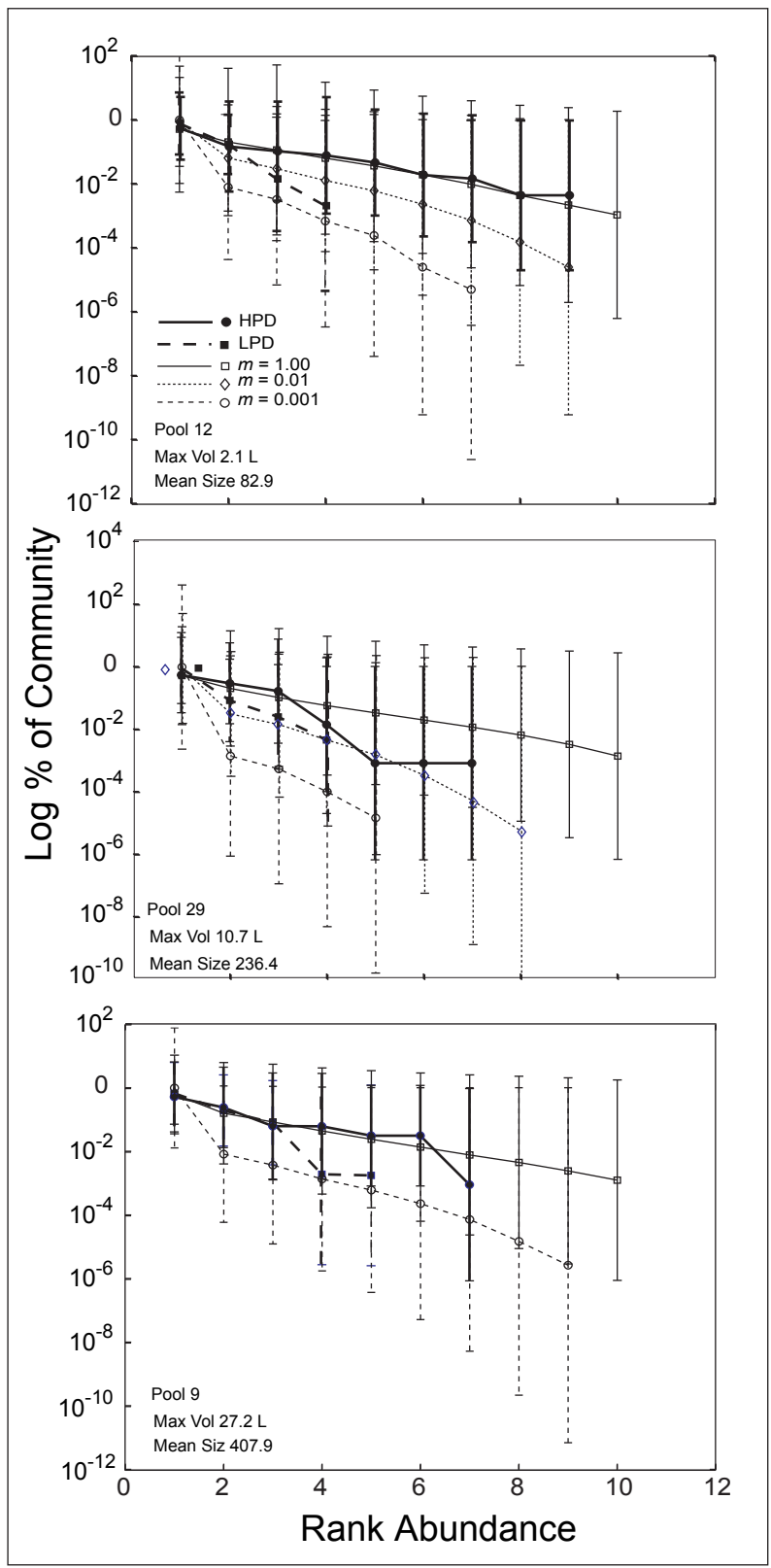

Figure 7. Predator Effect on Detritivore Communities

The ranked distributions of the logarithm of species average community proportions is shown for low predator density (LPD, closed squares, heavy dashed line) and high predator density (HPD, closed circles, heavy solid line) samples. Open symbols represent the logarithm of the proportions predicted for neutral communities of different migration rates, m. Error bars represent standard error. See figure for definitions of other symbols.